\documentclass[12pt,english]{article}
\usepackage[english]{babel}
\usepackage[dvips]{graphicx}
\makeatletter
\usepackage{amssymb}
\usepackage{latexsym}
\usepackage{amsmath}
\usepackage{bbm}
\usepackage{amsthm}

\newcommand{\RR}{\mathbbm{R}}    

\newcommand{\T}{\mathcal{T}}

\providecommand{\abs}[1]{\lvert#1 \rvert}       

\theoremstyle{definition}

\theoremstyle{remark}

\renewcommand{\T}{\mathbf T}
\newcommand{\x}{\mathbf x}
\renewcommand{\v}{\mathbf v}
\newcommand{\J}{\mathbf J}

\newcommand{\n}{\mathbf n}
\renewcommand{\RR}{\mathbb R}

\begin{document}
\title{Magnetworks: how mobility impacts the design of Mobile Networks}
\author{
Alonso Silva\footnote{INRIA Sophia-Antipolis,
France. Email: \texttt{\{alonso.silva, eitan.altman\}@sophia.inria.fr}}~$^{\dag}$,
Eitan Altman$^*$,\\
M\'erouane Debbah\footnote{
Alcatel-Lucent Chair in Flexible Radio - SUPELEC, France.
Email: \texttt{\{merouane.debbah, Giusi.Alfano@\}supelec.fr}},
Giuseppa Alfano$^{\dag}$
}

\maketitle

\begin{abstract}
In this paper we study 
the optimal placement and 
optimal number of active relay nodes through the traffic density in mobile 
sensor ad-hoc networks.
We consider a setting in which a 
set of mobile sensor sources
is creating data and a 
set of mobile sensor destinations receiving that data.
We make the assumption that the network
is massively dense, {\it i.e.}, there are so many sources, destinations,
and relay nodes, that it is best to describe the network in terms of
macroscopic parameters, such as their spatial density, 
rather than in terms of microscopic parameters, such as their individual
placements.

We focus on a particular physical layer model that is characterized
by the following assumptions: i)~the 
nodes must only transport the data from 
the sources to 
the destinations, and do not need to sense the data
at the sources, or deliver them at the destinations once the data arrive
at their physical locations, and ii) the nodes have limited
bandwidth available to them, but they use it optimally to locally
achieve the network capacity.

In this setting, the optimal distribution
of nodes induces a traffic density that resembles the electric
displacement that will be created if we substitute the sources
and destinations with positive and negative charges respectively. The
analogy between the two settings is very tight 
and have a direct interpretation in wireless
sensor networks.

\end{abstract}

\section{Introduction}


Various approaches inspired by physics have been proposed to deal
with the routing problem in massively dense wireless sensor networks. Starting with
the pioneering work of Jacquet (see~\cite{geometry,J2004}) who used ideas
from geometrical optics to deal with the case of one source and one
destination and a distribution of relay nodes.

Approaches based on electrostatics have been studied in
\cite{TT05,TT,GT,CMT} (see the survey~\cite{toumpis} and references therein)
to deal with the case of a distribution of sources and 
destinations with a density of 
relay nodes in a static environment.

The development of the theory of routing in massively dense wireless
sensor networks has emerged in a complete independent way of the
theory developed within the community of road traffic
engineers, introduced in 1952 by Wardrop~\cite{Wardrop} and
Beckmann~\cite{Beckmann}, see also~\cite[pp.~644, footnote
3]{BeckmannB} for the abundant literature of the early 50's, and which
is still an active research area among that community, see
\cite{Dafermos,daniele02,hoWong,idone04,wongsens} and references
therein.

Only very recently, new approaches based on road traffic theory have been studied
in~\cite{allerton,adhocnow,valuetools} defining
the Wardrop equilibrium and a characterization
of the Wardrop equilibrium in this type of setting
and including a geometrical characterization
of the flow of information for some particular cost functions.

Consider that a spatially distributed set of mobile sources
is creating data that must be delivered to a spatially distributed
set of mobile destinations. In this context, our objective is to study the
optimal traffic distribution and
to find the minimum amount of relay nodes needed to transport the data
from the sources to the destinations.

The main contribution of this work is to address this problem
for a mobile context where we analyze the cases where 
(i) only sources are mobile and the destinations are static as it
would be in the case when the aggregation centers are fixed
and the sensor nodes have the capability to move, (ii)
the case when both sources and destinations are mobile,
and given that the mathematics involve are similar (iii) we also
analyze the case when the sources are static and the
destinations are mobiles.

In section~\ref{one} we first analyze the case
where the sources and destinations are
static and spatially distributed in the one dimensional line
in order to illustrate the behavior of the flow function
and the optimal location of the relay nodes on this simple case.
We analyze in section~\ref{two} the case when the sources and
destinations are static and spatially distributed in the two dimensional plane.
In section~\ref{movement} we analyze the case
where the sources and destinations can move
with a deterministic velocity and we are able
to find the optimal flow of information and
to give the optimal spatial density
of the relay nodes at each period of time.
Within this section we also give an example with
numerical results related to the previous
mobility setting.
In section~\ref{brownian} we are able to
find the optimal spatial density of
the relay nodes for the Brownian mobility model.
In section~\ref{conclusions} we summarize
the main results and future perspectives for the continuation of our work.

\section{The Model}

We first consider the one dimensional case in order to explain the main
concepts involved in our model and how this concepts can be extended
to the two dimensional case in order to obtain the optimal deployment
of the relay nodes in a wireless sensor network.

\subsection{Fluid Approximations}
\label{one}

Consider a grid area network that contains wireless sensor nodes. As a first approach we
consider the line segment~$[0,L]$, which will be the geographical reference of a network.

We consider the continuous {\sl node density function}~$\eta(x)$,
measured in~$\textrm{nodes}/\textrm{m}$, 
%
such that the total number of nodes on a segment~$[\ell_0,\ell_1]$, denoted by $N(\ell_0,\ell_1)$,
is
\begin{equation*}
N(\ell_0,\ell_1)=\int\limits_{\ell_0}^{\ell_1} \eta(x)\,dx.
\end{equation*}
We consider as well the continuous {\sl information density function}~$\rho(x)$,
measured in $\textrm{bit}/(\textrm{s}\cdot\textrm{m})$,
generated by the sensor nodes
such that
\begin{itemize}
\item At locations~$x$ where $\rho(x)>0$ there is a fraction of
data created by the sensor sources, such that the rate with which information is created
in an infinitesimal area of size~$d\varepsilon$, centered at position~$x$, is equal to~$\rho(x)\,d\varepsilon$.
\item Similarly, at locations~$x$ where $\rho(x)<0$ there is a fraction of data received at the sensor destinations
such that the rate with which information is received by an infinitesimal area of size~$d\varepsilon$,
centered at position~$x$, is equal to~$-\rho(x)\,d\varepsilon$.
\end{itemize}

We assume that the total rate at which sensor destinations have to receive data is the same as the total rate which the data
is created at the sensor sources, {\it i.e.},
\begin{equation}\label{conserva}
\int\limits_0^L\rho(x)\,dx=0.
\end{equation}
Notice that if we have an estimation of the proportion of packet loss
in the network, we can ponderate our function~$\rho$ in order
to adequate it to equation~\eqref{conserva}.

Consider the continuously differentiable {\sl traffic flow function} $T(x)$, measured in $\textrm{bps}/\textrm{m}$, such that
its direction (positive or negative) coincides with the direction of the flow of information at point $x$ and
$\lvert T(x)\rvert$ is the rate at which information propagates at position $x$, {\it i.e.}, $\lvert T(x)\rvert$
gives the total amount of traffic that is passing through the position~$x$.

Next we present the flow conservation condition.
For information to be conserved over a segment~$[\ell_0,\ell_1]$,
it is necessary that the rate with which information is created over the segment, is equal to the
rate with which information is leaving the line segment, {\it i.e.},
\begin{equation*}
T(\ell_1)-T(\ell_0)=\int_{\ell_0}^{\ell_1}\rho(x)\,dx
\end{equation*}
The integral on the right hand side is equal to the quantity of information
generated (if it's positive) or demanded (if it's negative) by the fraction of sensor nodes
over the line segment~$[\ell_0,\ell_1]$.
The expression $T(\ell_1)-T(\ell_0)$, measured in $\textrm{bps}/\textrm{m}$,
is equal at the rate with which information is leaving (if it's positive) or
entering (if it's negative) the segment~$[\ell_0,\ell_1]$.
This holding for any line segment, it follows that necessarily,
\begin{equation}\label{conservation}
\frac{dT(x)}{dx}=\rho(x).
\end{equation}
The problem considered is to minimize the number of nodes~$N(0,L)$ in the line segment~$[0,L]$,
needed to support the information created by the sensor sources
and received by the sensor destinations
subject to the flow conservation condition given by equation~\eqref{conservation}
and imposing that there is no flow of information leaving the network, {\it i.e.},~$T(0)=0$ and~$T(L)=0$.
Thus the system of equations that model our problem in the one-dimensional case is given by the system of equations:
\begin{equation}
\operatorname*{Min} N(0,L)=\int_0^L\eta(x)\,dx,
\end{equation}
\begin{gather}
\text{subject to}\quad
\frac{dT(x)}{dx}=\rho(x)\quad\text{in} (0,L),\label{first}\\
T(0)=0\quad\text{and}\quad T(L)=0.\label{second}
\end{gather}
Notice that in the one-dimensional case, there is no minimization problem because only
by using the constraints~\eqref{first} and~\eqref{second}, we obtain only one solution. As we will see, this will not
be the case for the two-dimensional case.

We suppose that the proportion of sensor nodes~$\eta(x)$ in an area of infinitesimal size~$d\varepsilon$,
centered at location~$x$, needed as relay nodes,
will be proportional to the traffic flow of information that is passing
through that region, {\it i.e.},~$\eta(x)\,d\varepsilon=\lvert T(x)\rvert^{\alpha}\,d\varepsilon$
where~$\alpha>0$ is a fixed number called {\sl the relay-traffic constant}.
Then the optimal placement of the relay nodes in the network will be given by~$\eta^*(x)=\lvert T^*(x)\rvert^{\alpha}$,
where the traffic flow function~$T^*(x)$ is the optimal traffic flow function, given by the solution of the previous system of equations.
Furthermore, the optimal total number of relay nodes~$N^*(0,L)$ needed to support the optimal traffic flow function~$T^*(x)$
in the network will be
$
N^*(\ell_0,\ell_1)=\int_{\ell_0}^{\ell_1}\eta(x)\,dx=\int_{\ell_0}^{\ell_1}\lvert T(x)\rvert^{\alpha}\,dx.
$
Let us see an example to illustrate the previous framework.

{\it Example 1.-} Suppose that we can divide the line segment $[0,L]$ in two parts:
(i) in the first part~$[0,L/2]$ there will be a uniform information density function
generated by the sensor sources, given by \mbox{$\rho(x)=1$}
$\text{bps}/\text{m}$~and (ii) in the second half~$[L/2,L]$ there will be a uniform
information density function received at the sensor destinations given by
$\rho(x)=-1\text{~bps}/\text{m}$ (See Figure~1).
From the equations~\eqref{first} and~\eqref{second}
we obtain that the optimal traffic flow function will be given by~\mbox{$T^*(x)=x\text{~bps}/\text{m}$} for all $x\in[0,L/2]$
and~$T^*(x)=L-x\text{~bps}/\text{m}$ for all~$x\in[L/2,L]$ with positive direction (See Figure~2).
If we assume that the relay-traffic constant~$\alpha=2$, then the optimal placement of the relay nodes
needed to relay the information from the sources to the destinations
on the network will be given by~$\eta^*(x)=x^2$ for all~$x\in[0,L/2]$ and
$T^*(x)=(L-x)^2\text{~bps}/\text{m}$ for all~$x\in[L/2,L]$.

The optimal total number of relay nodes~$N^*(0,L)$ needed to support the optimal traffic~$T^*(x)$
will be given by
$
N(L)=\int_0^{L/2}x^2\,dx+\int_{L/2}^L(L-x)^2\,dx=L^3/12.
$

\begin{figure}
\centering
\includegraphics{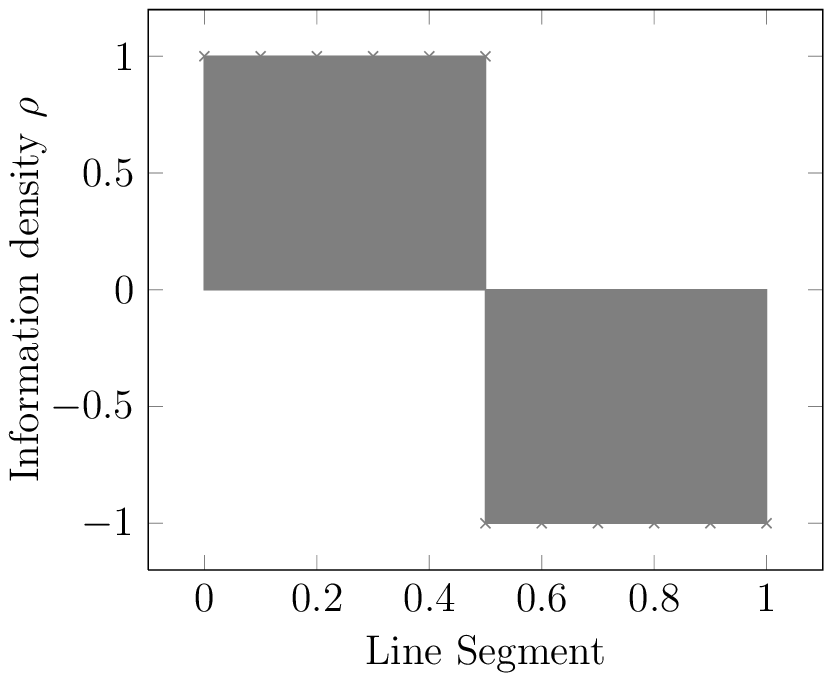}
\end{figure}

\begin{figure}
\centering
\includegraphics{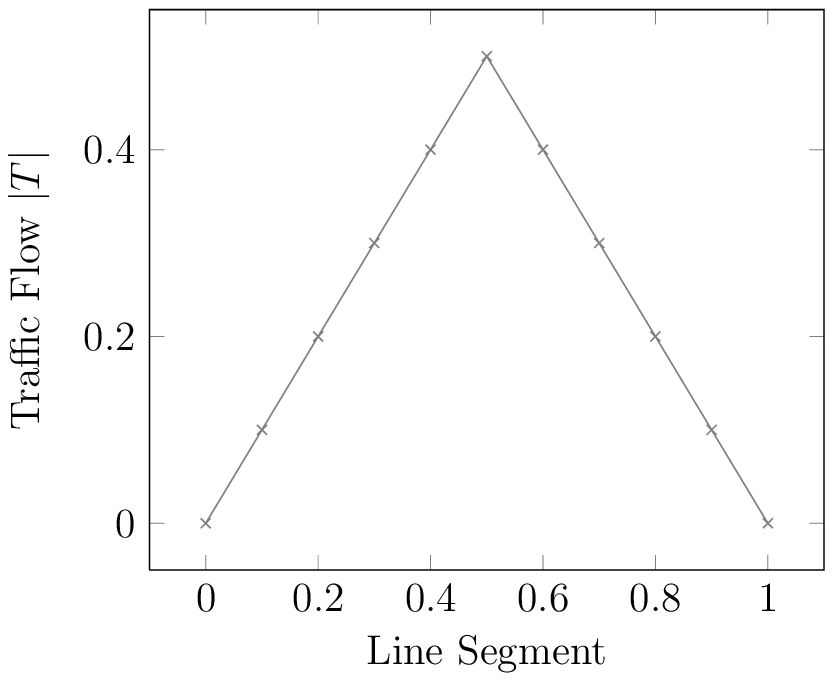}
\end{figure}

\begin{figure}
\centering
\includegraphics{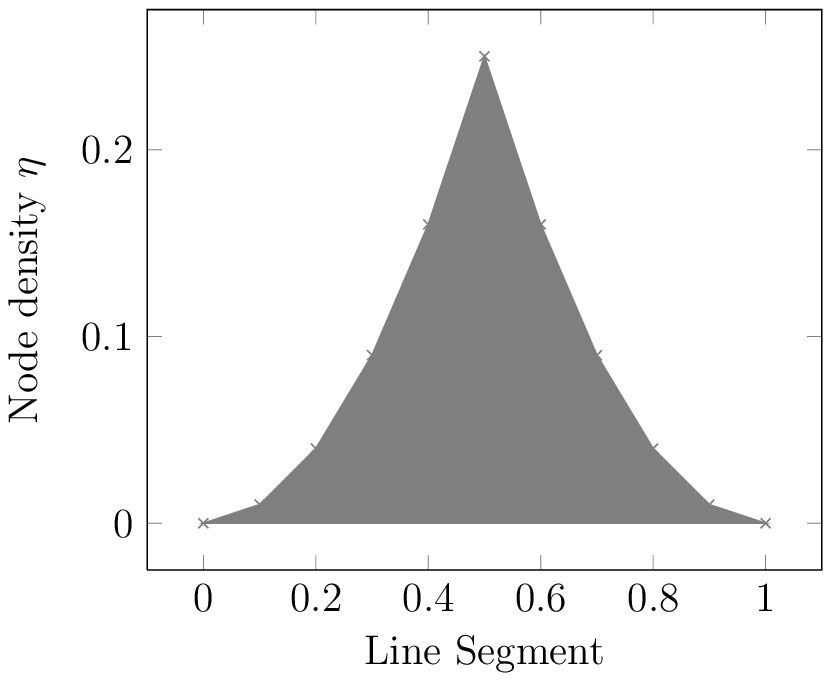}
\end{figure}

\subsection{The two dimensional case}
\label{two}

Consider a grid area network~$D$ in the two dimensional plane\footnote{We will denote with bold fonts the vectors and $\x=(x,y)$ will denote a
location in the two dimensional space~$X\times Y$.} $X\times Y$ the continuous 
{\sl information density function} $\rho(\x)$, measured in $\textrm{bps}/\textrm{m}^2$, such that at locations $\x$
where $\rho(\x)>0$, there is a distributed data created by sensor sources, such that the rate with which information is created
in an infinitesimal area of size $dA_\varepsilon$, centered at location~$\x$, is $\rho(\x)\,dA_\varepsilon$. 
Similarly, at locations $\x$ where $\rho(\x)<0$,
there is a distributed data received at sensor destinations,
such that the rate with which information can be treated by an infinitesimal area of size $dA_\varepsilon$,
centered at location~$\x$, is equal to~$-\rho(\x)\,dA_\varepsilon$.

The total rate at which sensor destinations must process data is the same as the total rate which the data
is created at the sensor sources, {\it i.e.},
\begin{equation*}
\int\limits_{X\times Y}\rho(\x)\,d\x=0.
\end{equation*}
Consider the continuous {\sl node density function}~$\eta(\x)$, measured in $\textrm{nodes}/\textrm{m}^2$,
defined so that the number of relay nodes in an area of infinitesimal size~$dA_\varepsilon$,
centered at $\x$, is equal to $\eta(\x)\,dA_\varepsilon$.

The total number of nodes on a region $A$, denoted by $N(A)$, is then given by
\begin{equation*}
N(A)=\int\limits_{A} \eta(\x)\,d\x.
\end{equation*}
Consider the continuous {\sl traffic flow function} $\T(\x)$, measured in $\textrm{bps}/\textrm{m}$, such that
its direction coincides with the direction of the flow of information at point $\x$, and\footnote{
The norm~$\lVert\cdot\rVert$ is the Euclidean norm, {\it i.e.}, for a vector~$\x=(x,y)$, its norm will be~$\lVert\x\rVert=\sqrt{x^2+y^2}$.}
$\lVert\T(\x)\rVert$ is the rate with which information rate crosses a linear segment perpendicular to $\T(\x)$
centered on $\x$, {\it i.e.}, $\lVert\T(\x)\rVert\,\varepsilon$ gives the total amount of traffic
crossing a linear segment of infinitesimal length~$\varepsilon$, centered at location~$\x$, and placed vertically~to~$\T(\x)$.

Next we present the flow conservation condition (see {\it e.g.}~\cite{TT, Dafermos}, for more details
about this type of condition).
For information to be conserved over a domain $D$ of arbitrary shape on the $X\times Y$ plane,
with smooth boundary $\partial D$,
it is necessary that the rate with which information is created in the area is equal to the
rate with which information is leaving the area, {\it i.e.},
\begin{equation*}
\int\limits_{D}\rho( \mathbf x )\,dD=
\oint\limits_{\partial D}[\mathbf T\cdot\mathbf{n}\,(\x)]\,d\ell
\end{equation*}
The integral on the left is the surface integral of~$\rho(\x)$ over the domain~$D$.
The integral on the right is the path integral of the inner product~$\T\cdot\mathbf n$
over the boundary $\partial D$. The vector $\mathbf n(\x)$ is the unit normal vector to $\partial D$
at the boundary point $\x\in\partial D$ and pointing outwards. 
Then the function $\T\cdot\mathbf n\,(\x)$, measured in $\textrm{bps}/\textrm{m}^2$, is equal at the rate
with which information is leaving the domain $D$ 
at the boundary point~$\x$.

This holding for any (smooth) domain~$D$, it follows that necessarily
\begin{equation}\label{conservate}
\nabla\cdot\T(\x):=
\frac{\partial {T}_1 (\x) }{ \partial x_1 } +
\frac{\partial {T}_2 (\x) }{ \partial x_2 } =
\rho(\x),
\end{equation}
where ``$\nabla\cdot$'' is the divergence operator.

Thus the problem considered is to minimize the quantity of nodes~$N(D)$ in the grid area network~$D$
needed to support the information created by the distribution of sources
subject to the flow conservation condition, {\it i.e.}, our problem is given by the system of equations:
\begin{gather}
\operatorname*{Min} N(D)\\
\text{subject to } \nabla\cdot\T=\rho(\x).
\end{gather}
Toumpis and Tassiulas in~\cite{TT05} focus on a particular physical layer model characterized by the following assumption:

{\em Assumption :} A location~$\x$ where the node density is~$\eta(\x)$ can support any traffic flow vector with a
magnitude less or equal to a bound~$\lVert\T(\x)\rVert_{\mathrm max}$ which is proportional to
the square root of the density, {\it i.e.} $\lVert\T(\x)\rVert\le\lVert\T(\x)\rVert_{\mathrm max}=K\sqrt{d(\x)}$.

The validity of Assumption~1 depends on the physical layer and the medium access control protocol used by the network.
Although it is not generally true, it holds in many different settings of interest.
For example, in~\cite{TT05} Toumpis and Tassiulas give an example of network where~$m^2$ nodes are placed
in a perfect square grid of $m\times m$ nodes and each node can listen to transmissions
from its four nearest neighbors. They give a simple time division routine so that the network of
$m^2$ nodes can support a traffic on the order of~$m$.
As another example, in~\cite{silvester} it was shown that the traffic that can be supported
in the above network, if nodes access the channel by use of slotted Aloha instead of time division,
is $\T_{\mathrm{local}}=K\times W\times m$, where nodes transmit data with a fixed global rate of $W$~\textrm{bps},
$K$ is a constant-smaller than $1/3$-that captures the efficiency of Aloha.
Finally, in~\cite{gupta99} it was shown that a network of~$n$ randomly placed nodes
can support an aggregate traffic on the order of $\sqrt{n/\log n}$
under a more realistic interference model that accounts for interference coming from
arbitrarily distant nodes. The logarithm in the denominator appears due to
the proving methodology of~\cite{gupta99}, and it has been shown~\cite{dousse}
that it can be dispensed off, by use of percolation theory in the proofs.

Tassiulas and Toumpis prove in \cite{TT05} that among all traffic flow functions that satisfy
$\nabla\cdot\T=\rho$, the one that minimizes
the number of nodes needed to support the network, must be irrotational, {\it i.e.},
\begin{equation}\label{curlfree}
\nabla\times\T=0.
\end{equation}
where ``$\nabla\times$'' is the curl operator.
\section{Model in motion}
\label{movement}
In our work we do a parallel to Electromagnetism by considering
moving sensor nodes or moving distribution of sensor nodes.
In that sense we first define some terms directly related to Electromagnetism.
In our model, the information density function~$\rho$, the traffic flow function~$\T$, and 
the node density function~$\eta$, defined previously may
depend on time, {\it i.e.}, 
$\rho=\rho(\x,t)$, 
$\T=\T(\x,t)$, and 
$\eta=\eta(\x,t)$.
We will consider our problem within a window of time~$t\in[t_i,t_f]$
where~$t_i$ is the initial time and~$t_f$ is the final time.

We define the continuous {\sl node current} $\J$ 
as the density of sensor nodes $\rho(\x,t)$
in the position $\x$ multiplied by the nodes average drift velocity
$\mathbf v (\x,t)$, {\it i.e.},
\[
\J=\rho(\x,t)\,\v(\x,t).
\]
The rate at which nodes leaves an area (or volume) $V$,
bounded by a curve (or surface) $S=\partial V$,
will be given by
\begin{equation}\label{left}
\oint_S \J\cdot\,dS
\end{equation}
Since the information density function is conserved in the plane this integral must be equal to
\begin{equation}\label{right}
\oint_S \J\cdot\,dS
=-\frac{d}{dt}\oint_{S}\rho\cdot\mathbf n\,dS
=-\int_V\frac{\partial\rho}{\partial t}\,dV.
\end{equation}
From the divergence theorem and imposing the equality between the equations \eqref{left} and \eqref{right},
we obtain the equivalent to Kirchhoff's current law:
\begin{equation*}
\nabla\cdot\J+\frac{\partial\rho}{\partial t}=0.
\end{equation*}
Notice as well that
\[
\nabla\cdot\J=\nabla\cdot(\rho\,\v)=\v\cdot\nabla\rho+\rho\nabla\cdot\v.
\]
We assume 
%
%
that we know the initial distribution of the sensor sources and the sensor destinations
at time~$0$ denoted by~$\rho_0$. Thus we obtain the following system of equations:
\begin{gather*}
\mathrm{(TE)}
\left\{
\begin{array}{rl}
\frac{\partial\rho}{\partial t}+\v\cdot\nabla\rho+\rho\nabla\cdot\v=0&\text{ in }D\times(0,T)\\
\rho(0)=\rho_0&\text{ on }D\times\{0\}.
\end{array}
\right.
\end{gather*}

The previous system of equations is known in the partial differential equations literature
as the linear transport equation with initial condition for which there exists a solution
(see Proposition~II.1 of~\cite{diperna}).

Notice that given the initial distribution of the sources and destinations and the velocity
of the distribution, we are able to compute the evolution of the distribution
of sensor sources and sensor destinations on time~$t\in[0,T)$. The velocity of sensor sources
and sensor destinations 
may be estimated by having some previous knowledge on the behavior of these sources and destinations
in our network.

We want to minimize the number of relay nodes~$N(D)$ in the grid area network~$D$
needed to support the information created by the distribution of sensor sources and
received by the distribution of sensor destinations subject to the flow conservation condition,
and knowing that the distribution of mobile sensor
nodes and mobile sensor destinations is the solution to the system of equations~$\mathrm{(TE)}$.
Thus our problem reads for all $t\in[0,T]$
\begin{equation}\label{finalcombat}
\operatorname*{Min}\,\,N(D,t)=\int_D \eta(\x,t)\,d\x=\\
=\int_D \lvert \T(\x,t)\rvert^{2}\,d\x
\end{equation}
\begin{subequations}
\begin{equation}
\text{subject to}\quad\nabla\cdot\T(\x,t)=\rho(\x,t)\text{ in }D,\label{conservationplus}
\end{equation}
\begin{equation}
\T\cdot\n=0\text{ on }\partial D.\label{nose}
\end{equation}
\end{subequations}
where~$\rho(\x,t)$ is the solution to the problem~$\mathrm{(TE)}$.

We recall that Tassiulas and Toumpis in~\cite{TT05} proved that among all
traffic flow functions that satisfy equation~\eqref{conservationplus},
the one that minimize the number of nodes needed to support the network,
must satisfy
\begin{equation}\label{roto}
\nabla\times\T=0.
\end{equation}
Using Helmholtz's theorem
(also known as fundamental theorem of vector calculus)
to last equation~\eqref{roto} we obtain that there exists a scalar potential function~$\varphi$
such that
\begin{equation}\label{scalarpotential}
-\nabla\varphi=\T.
\end{equation}
Replacing this function into the conservation equation~\eqref{conservationplus} we obtain that
\begin{equation}\label{laplace}
-\Delta\varphi=\rho
\end{equation}
and this holds for all~$t\in~[0,T)$.

We impose that no information is leaving the considered domain~$D$,
in equation~\eqref{nose} and from equation~\eqref{scalarpotential}
this condition translates into~$\nabla\varphi\cdot\n=0$

From equation~\eqref{laplace} and last condition
we obtain the following system
\begin{equation}
\mathrm{(LE)}
\left\{
\begin{array}{rl}
-\Delta\varphi=\rho&\text{ in }D\\
\nabla\varphi\cdot\n=0&\text{ on }\partial D. 
\end{array}
\right.
\end{equation}
which is the {\sl Laplace equation} with Neumann boundary conditions.

If the function~$f$ is square integrable then the Laplace equation
with Neumann boundary conditions has a unique solution in~$H^1(D)/\RR$.

In summary in order to solve our problem given by the equations
~\eqref{finalcombat},~\eqref{conservationplus},~\eqref{nose}, and $\mathrm{(TE)}$ 
we need to
\begin{enumerate}
\item\label{uno} Solve the system of equations~$\mathrm{(TE)}$, 
\item\label{dos} Put the solution as input into the system of equations~$\mathrm{(LE)}$, 
\item\label{tres} Solve the system of equations~$\mathrm{(LE)}$.
\end{enumerate}


Even as it looks complicated system we give an example where you can get explicit solutions.

%

%
%


{\sl Example 2.-}
Due to presentation effects we consider the one dimensional case during $T=2$ hours.\newline
We consider an initial distribution of sensor sources~$\rho_0^+$ and an initial distribution of sensor destinations~$\rho_0^-$
on the positive real line~$[0,+\infty)$, and we scale them to be probability distributions so it represents in each location
the proportion of sensor sources or the proportion of sensor destinations respectively:
\[
\rho_0^+=k_1 e^{-(x-3)^2}\quad\text{and}
\quad\rho_0^-=-k_2 e^{-(x-10)^2}.
\]
where~$k_1$ and $k_2$~ are normalization factors given by
$k_1=\frac{2}{\sqrt{\pi} {\rm erfc}(-3)}$,
$k_2=\frac{2}{\sqrt{\pi} {\rm erfc}(-10)}$,
where
${\rm erfc(x)}$ is the complementary error function defined as
\mbox{${\rm erfc(x)}=\frac{2}{\sqrt{\pi}}\int_x^{+\infty}e^{-s^2}\,ds$}.

We consider that the nodes average drift velocity is given by~$v(x,t)=x$.
We can think of a highway where the cars are equipped with sensors 
and while they are advancing on the highway they can go faster and faster.

\ref{uno}.- We first need to solve the transportation equation system~$\mathrm{(TE)}$, that in our example reads
\[
\left\{
\begin{array}{rl}
\frac{\partial\rho}{\partial t}+\frac{\partial(x\rho)}{\partial x}=0 & \text{on }\RR_+\times[0,T)\\
\rho(0)=\rho_0^+ +\rho_0^- & \text{on }\RR_+
\end{array}
\right.
\]

Using the method of characteristics we obtain that the 
information density function
over time is given by~$\rho(x,t)=\rho^+(x,t)+\rho^-(x,t)$ where
\[
\left\{
\begin{array}{l}
\rho^+(x,t)=k_1 e^{-(xe^{-t}-3)^2-t}\\
\rho^-(x,t)=-k_2 e^{-(xe^{-t}-10)^2-t}
\end{array}
\right.
\]
%

\begin{figure}
\centering
\includegraphics[width=8cm,clip]{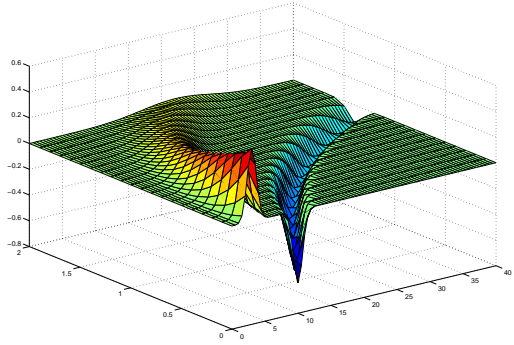}
\caption{Distribution of the sources and destinations in the same line}
\label{figsourcesdestinations}
\end{figure}

The solution combining the 
sources and destinations information density function 
over time are showed in Fig.~\ref{figsourcesdestinations}.


\ref{dos}.- We put the solution as input into the system of equations:
from the conservation equation we obtain
\[
\frac{\partial T(x,t)}{\partial x}=\frac{\partial T^+(x,t)}{\partial x}+\frac{\partial T^-(x,t)}{\partial x}
\]
\[
\text{where}\quad\frac{\partial T^+(x,t)}{\partial x}=\rho^+\quad\text{and}
\quad\frac{\partial T^-(x,t)}{\partial x}=-\rho^-.
\]
with initial condition that the flow is zero at the boundary point zero, {\it i.e.}~$T(0,t)=0$.

\ref{tres}.- We solve the laplacean system of equations:
Then the optimal traffic flow function is given by
\mbox{$T^{*}(x,t)=T^+(x,t)+T^-(x,t)$}
where
\[
\left\{
\begin{array}{l}
T^+(x,t)=\int_0^x k_1 e^{-(xe^{-t}-3)^2-t}\,dx\\
T^-(x,t)=-\int_0^x k_2 e^{-(xe^{-t}-10)^2-t}\,dx.
\end{array}
\right.
\]


Thus the minimal number of active relay nodes needed to support the optimal flow
at every time~$t$ will be given by
\[
N^*(t)=\int_0^{+\infty}\lvert T^{*}(x,t)\rvert^2\,dx
\]
which can be solved numerically.


\begin{figure}
\centering
\includegraphics[width=8cm,clip]{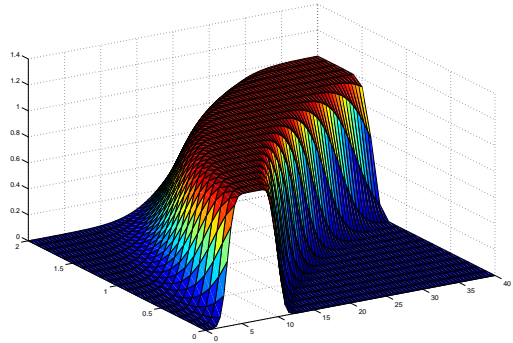}
\caption{Optimal traffic flow}
\end{figure}

\begin{figure}
\centering
\includegraphics[width=8cm,clip]{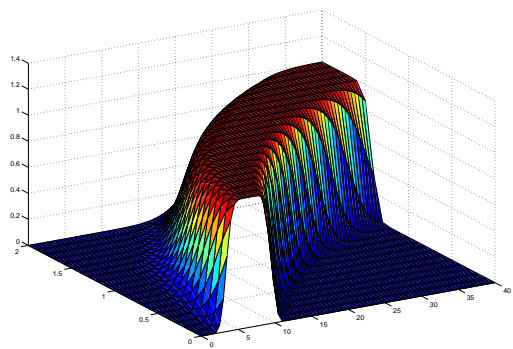}
\caption{Optimal relay node distribution}
\end{figure}

In next section we present another type of mobility model where we consider
the randomness in the mobility of the users.

\section{Brownian Mobility Model}
\label{brownian}

One of the most used mobility models used in networks is the
Random Walk Mobility Model also known as the Brownian Mobility Model (see the survey~\cite{surveymobility}
and the references therein).

If we have previous knowledge about the velocity drift of the distribution of information created at the sources
(denoted~$\rho^+$)
and/or the distribution of information received at the destinations (denoted~$\rho^-$),
and we assume the Brownian mobility model, then
the distribution of sources and/or the distribution of the destinations
evolves according to the stochastic differential equation
\[
d\rho^+(t)=\v^+(\x,t)\,dt+\sigma_+(\x,t)\,dW^+(t)
\]
\[
\text{and/or}\quad d\rho^-(t)=\v^-(\x,t)\,dt+\sigma_-(\x,t)\,dW^-(t).
\]
where~$W^+(t)$ and $W^-(t)$ are two independent
brownian motion with values in~$X\times~Y$ and
$\sigma_+:=\sigma_+(\x,t)$, $\sigma_-:=\sigma_-(\x,t)$ are parameters of the model. 

Assume as in the previous case that we know the initial distribution
of the information created at the sources.
Then by using It\^o's lemma, $\rho^+$ 
evolves in time
by the Kolmogorov Forward Equation
\[
\frac{\partial}{\partial s}p(\x,s)=-\frac{\partial}{\partial x}[\v^+(\x,s)p(\x,s)]
+\frac{1}{2}\frac{\partial^2}{\partial x^2}[\sigma_+^2 p(\x,s)].
\]
for~$s\ge 0$, with initial condition~$p(\x,0)=\rho^+(\x)$

Equivalently, the initial distribution of the destinations
evolves in time by the Kolmogorov Forward Equation
\[
\frac{\partial}{\partial s}p(\x,s)=-\frac{\partial}{\partial x}[\v^-(\x,s)p(\x,s)]
+\frac{1}{2}\frac{\partial^2}{\partial x^2}[\sigma_-^2 p(\x,s)].
\]
for~$s\in [t_i,t_f)$, with initial condition~$p(x,t_i)=\rho^-(x)$.

\section{Optimization on time}
Notice that the minimization problem we solved does not really consider the interaction on time 
because the problem describes the movement of sources and destinations nodes in the space
and then we solve the static problem at each time.
The problem solved at each time may not be optimal in the whole period of time considered.

Another more realistic problem would be to minimize the quantity of nodes used on the
whole network during a fixed period of time~$[t_i,t_f]$, {\it i.e.},
\begin{gather*}
\operatorname*{Min}\int_{t_i}^{t_f} N(D,t)\,dt=\int_{t_i}^{t_f}\!\!\!\int_D \eta(\x,t)\,d\x=\\
=\int_{t_i}^{t_f}\!\!\!\int_D\lvert\T(\x,t)\rvert^2\,d\x
\end{gather*}
where~$N(D,t)$ is the number of active relay nodes in the network~$D$ at time~$t$,
subject to~\eqref{conservationplus},~\eqref{nose}, and $\mathrm{(TE)}$. 

For the case we have randomness in the system this problem will be
\begin{equation}
\operatorname*{Min}\int_{t_i}^{t_f}\mathbb{E}\{N(D,t)\}\,dt=\int_{t_i}^{t_f}\!\!\!\int_D\mathbb{E}\left\{\lvert\T(\x,t)\rvert^2\right\}\,d\x
\end{equation}
subject to~\eqref{conservationplus},~\eqref{nose}, and $\mathrm{(TE)}$ 
since~$D$ is compact.
From the work of Santambrogio~(\cite{santambrogio}, page~6) 
we have the following result:
The problem 
\[
\operatorname*{Min}\int_D k(\x)\lvert\T(\x)\rvert\quad\textrm{such that}\quad\nabla\cdot\T=\mu-\nu
\]
is equivalent by duality to the problem of finding 
\[
\operatorname*{Min}\int_D d_k(x,y)\,d\gamma\quad\textrm{such that}\quad\gamma\in\Pi(\mu,\nu)\quad\text{where}
\]
\[
d_k(x,y)=\inf_{\{\omega\,:\,\omega(0)=x,\atop\quad\,\,\,\omega(1)=y\}}L_k(\omega):=\int_0^1 k(\omega(t))\lvert \omega'(t)\rvert\,dt
\]

In our case~$k(\x)=\lvert \T(\x)\rvert$ then
\[
L_k(\omega)=\int_0^1\lvert\T(\omega(t))\rvert\lvert\omega'(t)\rvert\,dt
\]
Given that $\omega(0)=x$, and $\omega(1)=y$ then by change of variables
$
L_k(\omega)=\int_x^y \lvert\T(\x)\rvert\,dx,
$
and as it is independent of~$\omega$ then $d_k(x,y)=\int_x^y \lvert\T(\x)\rvert\,d\x$.

{\it Example.-}
For the case where we do not have previous knowledge about the velocity drift
then we just consider the standard Brownian mobility model given by
\[
d\rho^+(t)=\sigma_+(\x,t)\,dW^+(t)
\]
and/or
\[
d\rho^-(t)=\sigma_{-}(\x,t)\,dW^-(t).
\]
where~$W^+(t)$ and $W^-(t)$ are two independent
Brownian motions with values in~$X\times~Y$.

Then the previous equations translate into
\[
\frac{\partial}{\partial s}p(x,s)=
+\frac{1}{2}\frac{\partial^2}{\partial x^2}[\sigma_{+}^2(x,s)p(x,s)].
\]

\[
\frac{\partial}{\partial s}p(x,s)=
+\frac{1}{2}\frac{\partial^2}{\partial x^2}[\sigma_{-}^2(x,s)p(x,s)].
\]

which have as solution the following equations
\[
\rho^+(x,t) = \frac{1}{\sqrt{2 \pi t \sigma^+}}e^{-\frac{x^2}{2t \sigma^+}}\,
\]

\[
\rho^-(x,t) = \frac{1}{\sqrt{2 \pi t \sigma^-}}e^{-\frac{x^2}{2t \sigma^-}}\,
\]

Now we can replace this solution into Step~\ref{dos} and Step~\ref{tres}.

{\it Remark.-}
Notice that if we suppose that the distribution of the destinations
is fixed, as it will be the case for aggregation centers of information,
then $\sigma^-=0$ and then $\rho^-(x,t)=\rho^-$ for all time~$t$.

\section{Conclusions}
\label{conclusions}

We consider a setting in which a spatially distributed set of mobile sources
is creating data for a spatially distributed set of mobile destinations.
We make the assumption that the network
is massively dense, {\it i.e.}, there are so many sources, destinations,
and nodes, that it is best to describe the network in terms of
macroscopic parameters, such as their spatial distribution, rather
than in terms of microscopic parameters, such as their individual
placements.
We focus on a particular physical layer model that is characterized
by the following assumptions: i) the wireless nodes
must only transport the data from the location of the sources
to the location of the destinations, and do not need to sense the data
at the sources, or deliver them at the destinations once the data arrive
at their physical locations, and ii) the nodes have limited
bandwidth available to them, but they use it optimally to locally
achieve the network capacity. 

In this setting, the optimal distribution
of nodes induces a traffic flow that resembles the electric
displacement that will be created if we substitute the sources
and destinations with positive and negative charges respectively. The
analogy between the two settings is very tight, and many features
of Electromagnetism have a direct interpretation in wireless
sensor networks.

\section*{Acknowledgement}

The authors want to thank Aim\'e Lachapelle from CEREMADE-Paris Dauphine for helpful discussions.
The first author was partially supported by CONICYT Chile and INRIA France.
The first, third and fourth authors were partially supported by Alcatel-Lucent
within the Alcatel-Lucent Chair in Flexible Radio at Supelec.
The mathematical results of section~\ref{brownian} are mainly taken from~\cite{lachapelle}.

\section{Appendix}

{\bf Definition.-}[Divergence]
Let $x, y, z$ be a system of Cartesian coordinates on a $3$-dimensional
space and let $\hat{\i},\hat{\j}$ and $\hat{k}$ be the corresponding basis of unit vectors respectively.
The {\bf divergence} of a continuously differentiable vector field $\mathbf F=F_x\,\hat{\i}+F_y\,\hat{\j}+F_z\,\hat{k}$
is defined to be the scalar-valued function:
\begin{equation*}
\nabla\cdot\mathbf F=\frac{\partial F_x}{\partial x}+\frac{\partial F_y}{\partial y}+\frac{\partial F_z}{\partial z}.
\end{equation*}

An equivalent definition is the following: Given a sequence of surfaces $A_k$, that all include in
their interior an arbitrary point $(x_0, y_0, z_0)$, such that their areas $\abs{A_k}\to 0$ with $k$,
then
\begin{equation*}
\nabla\cdot\mathbf F(x_0,y_0,z_0)=\lim_{k\to+\infty}\frac{1}{\abs{A_k}}\int_{\partial A_k}\mathbf F(x,y,z)\cdot\mathbf n\,dV,
\end{equation*}
where $\mathbf n(\x)$ is the unitary normal vector at $\x$.

We notice that both equivalent definitions can similarly be defined in a $2$-dimensional Euclidean space.

{\bf Definition.-}[Curl]
Let $x, y, z$ be a system of Cartesian coordinates on a $3$-dimensional
space and let $\hat{\i},\hat{\j}$ and $\hat{k}$ be the corresponding basis of unit vectors respectively.
The {\bf curl} of a continuosly differentiable vector field $\mathbf F=F_x\,\hat{\i}+F_y\,\hat{\j}+F_z\,\hat{k}$
is defined to be the vector field function:
\begin{gather*}
\nabla\times\mathbf F=
\left(\frac{\partial F_z}{\partial y}-\frac{\partial F_y}{\partial z}\right)\hat{\i}+
\left(\frac{\partial F_x}{\partial z}-\frac{\partial F_z}{\partial x}\right)\hat{\j}+\\
\left(\frac{\partial F_y}{\partial x}-\frac{\partial F_x}{\partial y}\right)\hat{k}.
\end{gather*}

{\bf Theorem.-} [Divergence Theorem]
The {\bf divergence theorem} states that for any well-behaved vector field $\mathbf A$ defined within
a volume $V$ surrounded by the closed surface $S$ the relation
\begin{equation*}
\oint_S \mathbf A\cdot\mathbf n\,dS=\int_V\nabla\cdot\mathbf A\,dV
\end{equation*}
holds between the volume integral of the divergence of $\mathbf A$ and the surface integral of the
outwardly directed normal component of $\mathbf A$.

The same result holds in a $2$-dimensional Euclidean space considering the corresponding definition of divergence.

{\bf Theorem.-}[Stokes Theorem]
The {\bf Stokes theorem} states that if $\mathbf A$ is a well-behaved vector field, $S$ is an arbitrary
open surface and $C$ is the closed curve bounding $S$, then
\begin{equation}
\oint_C\mathbf A\cdot d\ell= \int_S\left(\nabla\times\mathbf A\right)\cdot\mathbf n\,dS
\end{equation}
where $\mathbf n(\x)$ is the unitary normal vector at $\x$.

\end{document}